%Paper: astro-ph/9312061
%From: Vladimir Usov <FNUSOV@WEIZMANN.WEIZMANN.AC.IL>
%Date: Mon, 27 Dec 93 08:58:17 +0200

\magnification=1200  % Magnify everything by a factor of 1.2
\def\newline{\hfill\penalty -10000}  % Use to force a line break, if needed.
%==============================================================
\def\title #1{\centerline{\rm #1}}
\def\author #1; #2;{\line{} \centerline{#1}\smallskip\centerline{#2}}
\def\abstract #1{\line{} \centerline{ABSTRACT} \line{} #1}
\def\heading #1{\line{}\smallskip \goodbreak \centerline{#1} \line{}}
%================================================================
\newcount\refno \refno=1
\def\refjour #1#2#3#4#5{\noindent    \hangindent=1pc \hangafter=1
 \the\refno.~#1, #2 ${\bf #3}$, #4 (#5). \global\advance\refno by 1\par}
\def\refbookp #1#2#3#4#5{\noindent \hangindent=1pc \hangafter=1
 \the\refno.~#1, #2 (#3, #4), p.~#5.    \global\advance\refno by 1\par}
\def\refbook #1#2#3#4{\noindent      \hangindent=1pc \hangafter=1
 \the\refno.~#1, #2 (#3, #4).           \global\advance\refno by 1\par}
\newcount\equatno \equatno=1
\def\adveqn{(\the\equatno) \global\advance\equatno by 1}

\def\up#1{\leavevmode \raise 0.2ex\hbox{#1}}

%===========================================================================
% Definition of \figure: Includes advancing the counter (figno) at each call.
\newcount\figno
\figno=0
\def\figure{\global\advance\figno by 1 Figure~\the\figno.~}
%======================================================
%
% Remove ruled line separating footnotes from text

%
% Set up correct page dimensions
\vsize=8.75truein
\hsize=5.75truein
\hoffset=0.5truein
%
% Set baseline separation for 6 lines per inch
\baselineskip=0.166666truein
%
% Set paragraph indentation to the width of 5 average 10pt characters
\parindent=25pt
%
% Avoid skipping any extra space between paragraphs
\parskip=0pt
%
% Switch off page numbering
\nopagenumbers

%
%%%%%%%%%%%%%%%%%%%%%%%%%%%%%%%%%%%%%%%%%%%%%%%%%%%%%%%%%%%%%%%%%%%%%%%%%%%
%
% Leave a blank line before the main title
\line{}

\title{
% enter title text in capital letters on next line
GLITCHES IN THE X-RAY PULSAR 1E 2259+586
}

%%%%%%%%%%%%%%%%%%%%%%%%%%%%%%%%%%%%%%%%%%%%%%%%%%%%%%%%%%%%%%%%%%%%%
% enter authors name and address on the two lines after \author     %
% with a ; following each as shown                                  %
%%%%%%%%%%%%%%%%%%%%%%%%%%%%%%%%%%%%%%%%%%%%%%%%%%%%%%%%%%%%%%%%%%%%%
\author
Vladimir V. Usov \parindent=0pt ;
Dept.~of Physics, Weizmann Institute, Rehovot 76100, Israel ;

\abstract{
Starquakes are considered for fast-rotating magnetic white
dwarfs. The X-ray pulsar 1E 2259 + 586 may be such a white dwarf.
It is shown that in this case starquakes may be responsible for the
decrease of the mean spin-down rate which was observed for 1E 2259 + 586
between 1987 and 1990. The required mass of the white dwarf which
is identified with 1E 2259 + 586 is $\sim 1.4 - 1.5 M_\odot$,
making this X-ray pulsar the most massive white dwarf known.}

\heading{1. INTRODUCTION}
The X-ray pulsar 1E 2259+586 is located close to the center
of the young supernova remnant G$109.1 - 1.0$ whose age is
estimated to be $(1.2 - 1.7)\times 10^4$ yr (Gregory and Fahlman
1980; Hughes, Harten and van den Bergh
1981). The period of the pulsar, $P = 6.98$ s,
is suitable for an ordinary binary X-ray pulsar. But, 1E 2259 + 586
exhibits several peculiarities:
(1) the period shows an unusually stable spin-down trend with a small
rate of $5.9\times 10^{-13}$ s s$^{-1}$ (Davies, Wood and Coe 1990);
(2) no orbital Doppler modulation of the pulsar period has been
discovered (Koyama et al. 1989); (3) there is no firm evidence for
an optical counterpart (Davies and Coe 1991), and (4) it has an
unusually soft X-ray spectrum. Therefore, Davies et al. (1989) have
suggested that 1E 2259 + 586 is a single star. However, this X-ray pulsar
cannot be a single neutron star in which, like a radio pulsar, spin-down
power is a source of the radiation. The point is that for a neutron
star with the observed spin period and the spin-down rate, the
spin-down power cannot be larger than $10^{33}$ ergs s$^{-1}$,
while the soft X-ray luminosity of 1E 2259 + 586 is $\sim 10^{35}$
ergs s$^{-1}$ (Fahlman and Gregory 1981; Davies, Wood and
Coe 1990).
\par It was suggested by Paczy\'nski (1990) that the
enigmatic X-ray pulsar 1E 2259 + 586 is a single fast-rotating white
dwarf with a strong magnetic field, $B_{_S} \sim 10^9$ Gauss, at its
surface. The required mass of the white dwarf
is higher than $1.32 M_\odot$. This lower limit is because of a
centrifugal instability for the  white dwarf if its mass
is smaller than $1.32 M_\odot$. Such a massive white dwarf
may be the product of a white dwarf merger. Since the moment of
inertia for a massive white dwarf is more than $\sim 10^4$ times
larger than for a neutron star, the spin-down power of the white
dwarf may be as high as $\sim$ a few $\times 10^{36}$ ergs
s$^{-1}$, which is more than enough to explain the X-ray luminosity
of 1E 2259 + 586.
The soft X-ray emission of 1E 2259 + 586 can be explained by the
radiation of the hot gas in the polar caps (Usov 1993). This gas
is heated by the flux of positrons which are created in the
magnetosphere of the white dwarf.
\par It was shown recently (Iwasawa, Koyama and Halpern 1992) that
the spin-down rate, which was consistent with a constant value
for the first 10 years of observations, from 1978 to 1987,
has decreased by a factor
of 2 between 1987 and 1990. Using this, Iwasawa, Koyama and Halpern
(1992) have
concluded that the single white dwarf model of Paczy\'nski
(1990), which required a stable $\dot P$, is ruled out
for 1E 2259 + 586. In this paper I have argued that this
conclusion is premature. The point is that
if an essential part of the white dwarf is solid, starquakes have
to have happened occasionally in the process of
deceleration of the white dwarf rotation. In turn, these starquakes
may result in sudden changes of $P$, which are similar to the
pulsar glitches. One of these changes of $P$ may be responsible  for the
decrease of the mean spin-down rate which was observed between
1987 and 1990.

\heading{2. STARQUAKES AND THE PULSE PERIOD EVOLUTION}
Let us consider a fast-rotating magnetic white
dwarf with a solid core. As the rotation of the white dwarf
slows down because of the electromagnetic torque,
centrifugal forces on the core decrease, and gravity
pulls it toward a less oblate shape, thereby stressing it. When
stresses in the core reach a critical value, the core cracks,
some stress is relieved, and the excess oblateness, due to the core
rigidity, is reduced. The moment of inertia of the white dwarf is
suddenly decreased, and by conservation of angular momentum the
angular velocity of the white dwarf rotation, $\Omega =2\pi /P$,
is suddenly increased. This is similar to the starquakes which
were considered for neutron stars to explain the pulsar glitches
(Ruderman 1969; Baym and Pines 1971; Pines, Shaham and Ruderman 1972).
\par The oblateness of a rotating white dwarf may be characterized by
the following dimensionless  parameter:
$$\varepsilon ={{I - I_{_0}}\over I_{_0}}\,, \eqno (1)$$
where $I$ is the moment of inertia of the white dwarf, and $I_{_0}$ is
its moment of inertia in the case that the white dwarf does not rotate.
\par Since the angular momentum of the white dwarf is conserved in the
quake, the sudden change in the oblateness is
$$\Delta \varepsilon = {{\Delta I}\over I} = {{\Delta P}\over P}\,,
\eqno (2)$$
where $\Delta I$ is the change of $I$, and $\Delta P$ is the jump of
$P$.
\par The characteristic time between quakes is (Baym and Pines 1971)
$$t_q\simeq \tau _s{\omega _q^2 \over \Omega^2}\mid\Delta\varepsilon
\mid\,,\eqno (3)$$
where $\tau _s= P/{\dot P}$ is the time that characterized the rate at
which the white dwarf slows down due to loss of rotational energy
$\lbrack \tau _s$ is $\sim 3\times 10^5$ yr for 1E 2259 + 586 $\rbrack$,
$$\omega _q^2 ={2D^2 \over BI_{_0}}\,,\,\,\,\,\,\,\,\,\,\,\,\,\,
D={3\over 25}{GM^2_c\over R_c}\,, \eqno (4)$$
$$B=0.42 \left({4\pi \over 3}R^3_c\right){Z^2e^2n_{_A} \over a}\,,\,\,\,
\,\,\,\,\, a = \left({2\over n_{_A}}\right)^{1/3}\,, \eqno (5)$$
$n_{_A} =\rho _c/Am_p$ is the ion density, $\rho _c$ is the density of
matter in the white dwarf core,
$A$ is the atomic weight of ion, $Z$ is its electric charge,
$m_p$ is the mass of proton, $e$ is the charge of electron, $G$ is
the gravitation constant, $M_c$ is the mass of the white dwarf
core, and $R_c$ is its radius. Here and below we assume that the main
part of the white dwarf mass is solid. To get equations (4) and (5) for
$D$ and $B$
the solid core was considered as a rotating self-gravitating
sphere composed of incompressible matter of uniform shear modulus.
The estimate of $t_q$ by means of equation (3) is uncertain within
a factor of 2-3 or so.
\par The more reasonable parameters of the white dwarf which is
identified with 1E 2259 + 586 are the mass of the white dwarf $M\simeq
1.45 M_\odot$ (see below),
the stellar radius $R\simeq 3\times 10^8$ cm, $M_c
\simeq 1 M_\odot$, $R_c \simeq 2\times 10^8$ cm, $\bar \rho _c
=M_c/(4\pi /3)R^3_c \simeq 6\times 10^7$ g cm$^{-3}$, and
$I_{_0}\simeq 10^{50}$ g cm$^2$. Substituting these
parameters into equations (2) - (5), for the white dwarf which
consists of iron, $Z = 26$ and $A = 56$, we obtain $D\simeq 10^{50}$
ergs, $n_{_A} \simeq 7\times 10^{29}$ cm$^{-3}$, $a\simeq 1.4\times
10^{-10}$ cm, $B\simeq 10^{49}$ ergs, $\omega ^2_q \simeq 20$ s$^{-2}$,
$t_q\simeq 7\times 10^6
\mid \Delta \varepsilon\mid$ yr. Taking into account equation (2),
we have the following relation between $t_q$ and $\mid \Delta P
\mid /P$ for 1E 2259 + 586:
$$t_q\simeq 7\times 10^6{\mid\Delta P\mid \over P}\,\,\,{\rm yr}
\eqno (6)$$
\par The available observational
data on the changes of the pulse period of 1E 2259 + 586
from 1978 to 1990 are given in Figure 1. This data may be in agreement
with the single white dwarf model of 1E 2259 + 586 if we assume the
existence of glitches with $\Delta P = - (1-2)\times 10^{-5}$ s and
$t_q \simeq 3 - 10$ yr or more (see Fig. 1). For this value of $\Delta P$
equation (6) yields $t_q \simeq 10 - 20$ yr, which might be uncertain
within a factor of 2-3. This value of $t_q$ is consistent with
the data on the evolution of the pulse period of 1E 2259 + 586.
\par One of the main assumptions which we have used in this paper
is the existence of massive solid core inside the white dwarf which
is identified with 1E 2259 + 586. The cooling of white dwarfs and
their crystallization were considered mainly for white dwarfs with
mass $M\sim 0.6 M_\odot$, the "typical" mass remnant of single-star
evolution (for a review, see D'Antona and Mazzitelli 1990). For
such a white dwarf the time, $\tau _{cr}$, from its formation to the
stage of crystallization is $\sim 10^8 - 10^9$ yr. This is much more
than the age of the X-ray pulsar 1E 2259 + 586 in the model of
Paczy'nski (1990) in which the pulsar cannot be much
older than $\tau _s =P/\dot P \simeq 3\times 10^5$ yr, but it may be
as young as $\sim 1.5\times 10^4$ yr, if it is related to the supernova
remnant G109.1 - 1.0. The value of
$\tau _{cr}$ drops with increasing M, and it may be as
small as $\sim 10^5$ yr for the white dwarf with the mass
$M\simeq 1.4 - 1.5 M_\odot$ which relates to 1E 2259 + 586.
Indeed, for a plasma containing only one
species of ion, the temperature of crystallization is (see, for
example, Van Horn 1971; D'Antona and Mazzitelli 1990)
$$T_{cr} = 2.28\times 10^7 \Gamma ^{-1}{Z^2\over A^{1/3}}
\left({\rho \over 10^6\,\,{\rm g\,cm}^{-3}}\right)^{1/3}\,\,\,
{\rm K}\,, \eqno (7)$$
where $\Gamma$ is the dimensionless parameter which is somewhere
between 64 and 210. The more reasonable value of $\Gamma$ is 150.
\par Substituting $\Gamma = 150, Z =26$ and $A = 56$ into equation
(7) we have
$$T_{cr}\simeq 2.7\times 10^7\left({\rho \over 10^6\,\,{\rm g\, cm}
^{-3}}\right)^{1/3}\,\,\,{\rm K}\,. \eqno (8)$$
The temperature inside a massive white dwarf with the age $\sim
10^5$ yr is not higher than $\sim 10^8$ K (Vila 1966;
Savedoff, Van Horn and Vila 1969). A pure iron matter with this
temperature  has to crystallize if its density is equal or more than
$\sim 5\times 10^7$ g cm$^{-3}$ (see equation (8)). Both the mean
density and the density at the center of a white dwarf increase
very sharply when the white dwarf mass goes to the edge of stability
(Geroyannis and Hadjopoulos 1989 and references therein).
Using the paper of Geroyannis and Hadjopoulos (1989), we can see that
the main part of the white dwarf mass is inside of the surface at
which the density is equal to $5\times 10^7$ g cm$^{-3}$ only if
the mass of the white dwarf is $\sim 1.4-1.5 M_\odot$.
Therefore, if the white dwarf, which is identified with 1E 2259 + 586,
consists of iron and its mass is near the edge of stability,
$M\simeq 1.4 - 1.5 M_\odot$, the assumption that
the main part of the white dwarf mass is solid is reasonable.

\heading{3. DISCUSSION}
In this paper I have suggested the existence of
glitches with the amplitude $\Delta P\simeq - (1-2)\times 10^{-5}$ s
and the characteristic time between glitches $t_q \simeq$ a few
$\times (1-10)$ yr for the enigmatic X-ray pulsar 1E 2259 + 586.
This is in agreement with the single
white dwarf model of Paczy\'nski (1990) for 1E 2259 + 586.
The observation of glitches would be a confirmation of this model.
\par The other way to verify the single white dwarf
model is the observation of
$\gamma$-rays from 1E 2259 + 586. The point is that fast-rotating
white dwarfs, $P\simeq$ a few s, with strong magnetic field,
$B_{_S}\sim 10^8-10^9$ Gauss, are similar to the radio pulsars, and
both the particle acceleration to ultrarelativistic energies and
generation of $\gamma$-rays have to be in their magnetospheres
(Usov 1988, 1993). If 1E 2259 + 586 is such a white dwarf, the
expected flux of $\gamma$-rays with energies $\sim 10^2 - 10^3$
MeV may be high enough to be detected by EGRET (Usov 1993).

\heading{FIGURE CAPTION}
{\it Figure 1.} Pulse period histiry of 1E 2259 + 586
from 1978 to 1990. The solid line is the best fit to the points
before 1989 without glitches. The dashed lines show the
fit to the points with glitches. The amplitude of the glitches is
somewhere between $10^{-5}$ s and $2\times 10^{-5}$ s. The
characteristic time between glitches may be as small as a few years
(see Fig. 1b). In this case the spin-down rate between glitches is
essentially more than the mean spin-down rate for the first 10 years
of observations.

\heading{ACKNOWLEDGMENTS}
I thank J.P. Halpern for sending his paper which stimulated this work
and Harry Shipman, the referee, for many helpful suggestions
that improved the final manuscript.

\heading{REFERENCES}
\noindent
Baym, G., and Pines, D. 1971, Annals of Physics, {\bf 66}, 816

\noindent
D'Antona, F., and Mazzitelli, I. 1990, ARAA, {\bf 28}, 139

\noindent
Davies, S.R., and Coe, M.J. 1991, MNRAS, {\bf 249}, 313

\noindent
Davies, S.R., Coe, M.J., Payne, B.J., and Hanson, C.G. 1989,
MNRAS, {\bf 237}, 973

\noindent
Davies, S.R., Wood, K.S., and Coe, M.J. 1990, MNRAS, {\bf 245}, 268

\noindent
Fahlman, G.G., and Gregory, P.C. 1981, Nature, {\bf 293}, 202

\noindent
Geroyannis, V.S., and Hadjopoulos, A.A. 1989, ApJS, {\bf 70},
661

\noindent
Gregory, P.C., and Fahlman, G.G. 1980, Nature, {\bf 287}, 805

\noindent
Hughes, V.A., Harten, R.H., and van den Bergh,S. 1981, ApJ,
{\bf 246}, L127

\noindent
Iwasawa, K., Koyama, K., and Halpern, J.P. 1992,
PASJ, {\bf 44}, 9

\noindent
Koyama, K. et al. 1989, PASJ, {\bf 41}, 461

\noindent
Paczy\'nski, B. 1990, ApJ, {\bf 365}, L9

\noindent
Pines, D., Shaham, J., and Ruderman, M.A. 1972, Nature Phys. Sci.,
{\bf 237}, 83

\noindent
Ruderman, M.A. 1969, Nature, {\bf 223}, 597

\noindent
Savedoff, M.P., Van Horn, H.M., and Vila, S.C. 1969, ApJ,
{\bf 155}, 221

\noindent
Usov, V.V. 1988, Sov. Astron. Lett., {\bf 14}, 258

\noindent
Usov, V.V. 1993, ApJ, {\bf 410}, 761

\noindent
Van Horn, H.M. 1971, in {\it IAU Symposium 42, White Dwarfs,}
(Dordrecht: Reidel), p. 97

\noindent
Vila, S.C. 1966, ApJ, {\bf 146}, 437

\end